%% file: main.tex
\documentclass{article}

\input{packages.tex}

\input{definitions.tex}

\title{Blind Deinterleaving of Signals in Time Series with Self-attention Based Soft Min-cost Flow Learning}

\input{authors.tex}

\begin{document}
%
\maketitle
\begin{abstract}
We propose an end-to-end learning approach to address deinterleaving of patterns in time series, in particular, radar signals. We link signal clustering problem to min-cost flow as an equivalent problem once the proper costs exist. We formulate a bi-level optimization problem involving min-cost flow as a sub-problem to learn such costs from the supervised training data. We then approximate the lower level optimization problem by self-attention based neural networks and provide a trainable framework that clusters the patterns in the input as the distinct flows. We evaluate our method with extensive experiments on a large dataset with several challenging scenarios to show the efficiency.
\end{abstract}
\section{Introduction}
\label{sec:intro}

Practical Electronic Warfare (EW) systems are often operated in environments where multiple signal sources exist simultaneously. Signal patterns from different emitters are received via the same channel of the receiver in time order, hence obtaining an interleaved signal sequence. For robust classification of radar signals, obtaining the individual sequences from emitters (\ie deinterleaving, emitter separation or signal sorting) is essential \cite{wiley2006elint, barton2004radar}. 

The first degree attributes of radar pulse streams are time of arrival (ToA), direction of arrival (DoA), pulse amplitude (PA), pulse width (PW) and radio frequency (RF), which are stored in pulse description words (PDWs) \cite{wiley2006elint, barton2004radar}. DoA, PA, PW and RF may be missing or follow unpredictable irregular patterns \cite{mardia1989new, milojevic1992improved} though clustering the pulses with respect those attributes seem to be appealing. On the other hand, radar signals usually have characteristic pulse repetition interval (PRI) which is the first difference of the ToA sequence. Following the existing efforts \cite{mardia1989new, milojevic1992improved, nishiguchi2000improved, 9123957}, we exploit only ToA measures to deinterleave radar signals .  

Prior conventional methods construct histograms from ToA differences to capture the periodicity of the pulses \cite{mardia1989new, milojevic1992improved}. While the cumulative difference histogram (CDIF) \cite{mardia1989new} accumulates histogram values for each difference level of the ToA sequence in a single iteration, the sequential difference histogram (SDIF) \cite{milojevic1992improved} computes the histogram values simultaneously for each difference order. Following that, they estimate PRI values of radar signals from calculated histogram by using heuristic threshold functions and sequentially search and cluster pulses, which follows estimated PRI regimes to form radar signals. Since those approaches are sensitive for sub-harmonics, PRI transform (PRIT) \cite{nishiguchi2000improved} proposes to use a complex-valued autocorrelation-like integral instead of calculating histograms with the assumption that the ToA sequence is quite large. Although those conventional algorithms work well for simple PRI modulation types, they fail to succeed in scenarios where largely missing pulses, highly jittered PRI values and staggered PRI modulation modes exist. 
\input{figures/fig_the_figure_1.tex}

Recently, heuristic approaches \cite{ge2019improved, liu2018improved, mahdavi2011fast, mao2009improved} are proposed to deinterleave radar signals which have complex PRI modulation types. However, their limitation to specific scenarios prevents generalizing to problems of different nature. Furthermore, machine learning techniques have been applied to solve the deinterleaving task. For instance, fuzzy adaptive resonance theory (ART) is utilized in \cite{ata2007deinterleaving} to cluster pulse streams by exploiting RF, PW and DoA features before using one of the histogram based deinterleaving methods. Moreover, \cite{liu2018classification} utilizes recurrent neural networks (RNNs) to capture the current context of the pulse stream and predict attributes of the next PDWs. Yet, RNNs classify PDWs from predefined radar list, hence, it identifies pulses rather than solving the deinterleaving problem. Finally, denoising pulse streams with autoencoders \cite{li2020deinterleaving} is visited to improve the deinterleaving solution.

We address alleviating the limitations of the aforementioned approaches through a novel end-to-end trainable clustering framework. Starting from linking deinterleaving to min-cost flow problem, we develop a bi-level optimization problem and relate the solver of the lower problem to the residual networks, yielding a novel unified self-attention based framework enabling to learn clustering arbitrary number of patterns in signals.

\section{Method}
\label{sec:method}
\subsection{Min-cost Flow Formulation of Signal Clustering}
We consider deinterleaving of PRI patterns using relative time of arrival (RToA) measures though our formulation is general for deinterleaving in time series. RToA is computed from ToA differences: $\text{RToA}_i {=} \text{ToA}_i {-} \text{ToA}_{i{-}1}$  for $i{>}1$ with $\text{RToA}_1 {=} 0$.

We have $N$ RToAs, $x{=}\lbrace x_i\rbrace_i{\in} \R^{N}$, of mixed signals. Our aim is to divide $x$ into $K$ disjoint sets, $x^{(k)}{=}\lbrace x_{k_i}\rbrace_i{\in} \R^{N_k}$, containing the RToAs of the distinct sources. We encode such sets by the ground truth assignment matrix, $y{=}\lbrace y_{ij}\rbrace_{ij}\in\lbrace 0,1 \rbrace^{N{\times}N}$, where $y_{ij}{=}1$ tells $x_i$ is preceding $x_j$ in a distinct pattern, \ie $x_i{=}x_{k_i}$ and $x_j{=}x_{k_{i+1}}$ as  $x^{(k)}{=}\lbrace\ldots x_i, x_j \ldots\rbrace$.

We formulate deinterleaving problem as a min-cost flow problem on a directed graph. The vertices of the graph are RToAs which we denote as $x{=}\lbrace x_i\rbrace_i\in \R^{N}$ and the edges are the possible connections among RToAs. We augment the graph with a start and a terminal vertex, $x\cup\lbrace x_s,x_t\rbrace$, to help formulation. We associate edges with costs, $c=\lbrace c_{ij} \rbrace_{ij}\in\R^{(N{+}1){\times}(N{+}1)}$, and mean to find min-cost flows from some starting vertices to the terminal vertices. Such flows will be the deinterleaved signals. Formally, we consider the problem:
\begin{equation} \label{eq:mincost_flow}
    p^{\ast} =\argmin_{\substack{
    \sum_j p_{ij}=1, \forall i\setminus\lbrace s \rbrace\\
    \sum_i p_{ij}=1, \forall j\setminus\lbrace t \rbrace\\
    \sum_{j} p_{sj}=\sum_{i} p_{it}\\
    0\leqslant p_{ij}\leqslant 1, \forall ij}}
    \textstyle\sum_{ij}c_{ij}\, p_{ij} \tag{P1}
\end{equation}
to find an assignment matrix, $p$, from which we can recover the flows. The constraints avoid one-to-many or many-to-one connections (1\&2) and preserve flow continuity (3). \textit{Total unimodularity} of the constraints ensures integer solutions for $p_{ij}$ \cite{lpbook}, meaning that we have a binary assignment matrix, $p$, which encodes the linking of the flows. We note that the number of sources, $K$, can be arbitrary; since, no explicit dependency on $K$ exists in the formulation.

\subsection{Learning Association Costs and Soft Min-cost Flow}
\label{sec:learning}
The optimality of the computed flows are based on the edge costs, $c_{ij}$. Ideally, given $y$ as the ground truth assignments, $c{=}1{-}y$ would yield the flows as the desired deinterleaved signals. In practice, we have no information of $y$.

We are to learn association costs from some training data, $x$, with known assignments, $y$. To formulate our learning framework, we express the cost matrix as a parametric function of $x$ with parameters $\theta$ to be learned as $c(\theta)$ where we drop the dependency on $x$ for simplicity. Then the problem becomes a bi-level optimization problem:
\begin{equation} \tag{P2} \label{eq:learning_costs}
\begin{split}
    \theta^\ast &= \argmin -\tfrac{1}{N}\textstyle\sum_{ij}y_{ij}\log p^\ast_{ij}\\
    &\text{subject to } p^\ast=\argmin_{\substack{A\,p=b\\
    0\leqslant p_{ij}\leqslant 1, \forall ij}} c(\theta)\T p
\end{split}
\end{equation}
where $A$ and $b$ are the flow constraints in \eqref{eq:mincost_flow}. The Jacobian of the linear program (LP) $\tfrac{\partial p^\ast}{\partial c}$ must be computed to perform gradient updates for $\theta$. This can be achieved by transforming LP into an unconstraint problem with unique solution and using implicit differentiation \cite{ochs2015bilevel}. Equality constraints can be eliminated by expressing $p{=}p_0{+}Bz$ where $p_0$ is any solution to $A\,p_0{=}b$ and $B$ is the basis for the null space of the constraints $A$. The inequality constraints can be relaxed by introducing strictly convex penalty terms as in \cite{schulter2017deep}. Hence, the relaxed unconstrained problem, $z^\ast{=}\argmin\mathcal{L}(z)$, has a unique solution to be obtained by gradient updates over $z$ as:
\begin{equation}\label{eq:grad_update}
    z^{(l{+}1)}= z^{(l)} - \alpha\,\grad_z\,\mathcal{L}(z)\mid_{z{=}z^{(l)}}
\end{equation}
where $l$ is the iteration number. For a solution obtained at the $L^{th}$ iteration, $z^{(L)}$,  computing the partial derivatives for $\tfrac{\partial p^\ast}{\partial c}$ can be seen as applying chain rule recursively on $z^{(l)}$s: $\tfrac{\partial z^{(1)}}{\partial c}\prod_{l{=}1}^L \tfrac{\partial z^{(l)}}{\partial z^{(l{-}1)}}$, where we deliberately omit dependency on $c$ for simplicity. To this end, with the abuse of the gradient expression, we consider each update as a non-linear transform of $z^{(l)}$ added to itself:
\begin{equation}\label{eq:grad_residual}
z^{(l{+}1)}= z^{(l)} + g(z^{(l)})  
\end{equation}
which is similar to residual connections in deep learning \cite{resnet, vaswani2017attention}. The expression in Eqn. \eqref{eq:grad_residual} informally suggests approximating a solver by stacked residual layers. We use this resemblance to build our formulation which couples cost learning and LP solving within a single  framework. Specifically, we consider assignment matrix as a parametric function of $x$ directly, $p(x;\theta)$, rather than the solution of an LP and formulate the learning problem as:
\begin{align}
    \theta^\ast &= \argmin_\theta -\tfrac{1}{N}\textstyle\sum_{ij}y_{ij}\log p(\theta)_{ij} \tag{P3} \label{eq:soft_network_flow}\\
    \text{subject to } &\textstyle\sum_j p(\theta)_{ij}\leqslant 1, \forall i \label{eq:row_const}\tag{C3.1}\\
    &\textstyle\sum_i p(\theta)_{ij}\leqslant1, \forall j{\setminus}\lbrace t \rbrace \label{eq:col_const}\tag{C3.2}\\
    &\textstyle\sum_i p(\theta)_{it}=\textstyle\sum_{j} (1{-}\textstyle\sum_i p(\theta)_{ij})\label{eq:flow_cont}\tag{C3.3}\\
    &p(\theta)_{ij}\in\lbrace 0,1\rbrace, \forall ij \label{eq:binary_const}\tag{C3.4}
\end{align}
where $p(x;\theta)$ is realized by a stack of \textit{bidirectional transformer encoder} \cite{vaswani2017attention} layers. We omit start node, $s$, and related constraints from the formulation and embed flow continuity in \eqref{eq:flow_cont}. We explicitly introduce flow constraints to the learning problem in terms of $p$. Thus, the constraints are no longer linear nor convex. However, such a formulation allows us to use supervision to learn solving min-cost flow problem with proper costs inherently. Hence, no explicit LP solver to compute assignment matrix is required in practice.

We solve \eqref{eq:soft_network_flow} by relaxing the constraints with proper penalty terms to obtain an unconstrained problem. As we will show in Sec. \ref{sec:arch}, \eqref{eq:row_const} is already satisfied by design. We use hinge loss for \eqref{eq:col_const},
\begin{equation}
    \mathcal{L}_2(\theta)=\tfrac{1}{N}\textstyle\sum_j\operatorname{max}(0,\textstyle\sum_i p(\theta)_{ij}-1)
\end{equation}
and squared L2 norm penalty for \eqref{eq:flow_cont},
\begin{equation}
    \mathcal{L}_3(\theta)=\Vert \textstyle\sum_i p(\theta)_{it}{-}\textstyle\sum_{j} (1-\textstyle\sum_i p(\theta)_{ij}) \Vert_2^2\,.
\end{equation}
For the binary constraint \eqref{eq:binary_const}, we exploit the fact $\Vert p_i \Vert_1\geqslant \Vert p_i \Vert_2$ always and equality holds when $p_i$ is a one-hot vector.
\begin{equation}
    \mathcal{L}_4(\theta) = \tfrac{1}{N}\textstyle\sum_i(\Vert p_i \Vert_1 - \Vert p_i \Vert_2)
\end{equation}
where $p_i$ is the i$^{th}$ row of $p(\theta)$. Hence, total loss becomes:
\begin{equation}\label{eq:final_loss}
\begin{split}
    \mathcal{L}_{flow}(\theta) = &-\tfrac{1}{N}\textstyle\sum_{ij}y_{ij}\log p(\theta)_{ij} \\ &+\lambda_2\,\mathcal{L}_2(\theta) +\lambda_3\,\mathcal{L}_3(\theta)+\lambda_4\,\mathcal{L}_4(\theta)
\end{split}
\end{equation}
where $\lambda_i$s are the penalty coefficient hyperparameters. Once the parameters, $\theta$ are learned by $\theta^\ast{=}\argmin\mathcal{L}_{flow}(\theta)$, distinct patterns in a sequence, $x$, can be clustered using computed flows.

\subsection{Soft Min-cost Flow Architecture: p(x;$\vect{\theta}$)}
\label{sec:arch}
We exploit stack of \textit{bidirectional transformer encoder} (BTE) \cite{vaswani2017attention} layers to model $p(x;\theta)$ owing to the success of the architectures \cite{devlin2018bert, vaswani2017attention} equipped with such layers in encoding the high order relations among sequential data. A BTE layer is a stack of a self-attention and a feed-forward layer each of which contains a non-linear function and a residual connection. Thus, such layers meet our intuition for expressing a solver in terms of stacked residual units as in Eqn. \eqref{eq:grad_residual}.

Self-attention layers expresses the representation of each sample in the sequence as the convex combination of the representation of the all samples. The mixing weights are according to the relative similarities of the representations. Thus, stack of such layers hierarchically encodes higher order relations among the samples of the sequential data.

In our architecture which is depicted in Fig. \ref{fig:the_figure_1}, we first represent each sample, $x_i$, in the sequence with a $D$-dimensional vector. We normalize the input sequence, $\bar{x}{=}\tfrac{x}{\operatorname{max}_i\lbrace x_i \rbrace_i}$, and then assign each sample, $x_i$, to some token with embedding $\in\R^D$ according to quantization of the interval, $[0,1]$. One can also use a neural network to embed $x_i$s to $\R^D$; however, we empirically found that the former works well in practice and simpler. We then perform relative position encoding as in \cite{shaw2018self} to emphasize ordering in time. These encoded representations are then fed to the stack of $L$ BTE layers to yield transformed representations, $\lbrace r_i\rbrace_i$, which are enhanced with the higher order relations. We apply affine transformation followed by soft-max to these representations: $q_i = T\,r_i + \beta$ where we consider columns of $T$ as the decision representatives which tell to which the corresponding sample is to be linked . Thus, $q_{ij}$ is a proxy of the similarity between $r_i$ and the $j^{th}$ decision representative. Large $q_{ij}$ means $i$ is probably linked to $j$. Finally, soft-max of $q$ reads the assignment matrix and note that the constraint \eqref{eq:row_const} in \eqref{eq:soft_network_flow} is readily satisfied with this formulation: $ p_{ij}=q_{ij}/\textstyle\sum_jq_{ij} $.

Though using fixed decision representatives brings the limitation of processing fixed-length sequence, this limitation can be alleviated during inference by processing the sequences in overlapping windows. We empirically found that interaction of the samples too far from each other can be neglected. Thus, fixed-length training of appropriate size achieves good generalization performance.

\subsection{Inference}
\label{sec:inference}
In general, we have soft assignments in our estimated assignment matrix, $p$. We compute a binary assignment matrix from $p$ by solving \eqref{eq:mincost_flow} with $1{-}p$ as the cost matrix. We also experimented greedily performing assignments starting from the highest $p_{ij}$. We provide the comparisons in Sec. \ref{sec:qres}.

For the sequences whose lengths exceed the model capacity, we take overlapping segments of the sequence and perform assignments. We take the most recent assignment for a sample in case of multiple assignments due to overlapping.

\section{Experimental Work}
\label{sec:experiment}

\input{figures/table_performance}

\subsection{Datasets}
We use simulator to generate ToA sequences for training and validation data. PRI types of the radars (Fig. \ref{fig:pri_types}) are constant, jitter, constant stagger, random stagger and switch\&dwell \cite{ahmadi2012pri}. To produce realistic data, we select the maximum standard deviation of constant, jitter, constant stagger, random stagger and switch\&dwell for ToA are +/-1\%,  +/-15\%,  +/-40\%, +/-40\% and +/-40\%, respectively. Since normalization occurs in PRI as pre-processing of the algorithm, we choose PRI values from pre-defined range which is between 1 and 1000 $\mu$s. Stagger level can vary from 2 to 9. We sample switch\&dwell repetition parameter (the number of dwells) between 4 and 10 where the range of the number of switches is kept as same as the stagger level range. The least and the most number of sample belong to same radar is 5 and 100, respectively. We set the maximum number and minimum of radars in a sequence is 10 and 1, respectively.

We generate 200000 interleaved radar sequences for training and 5000 for validation data. Sequences in each dataset for the following cases are sampled with equal probability: \textbf{Case 1:} Only constant and jitter PRI type. \textbf{Case 2:} No missing occurs. While ending times of radar sequences are same, beginning times of radar sequences are close. \textbf{Case 3:} Missing occurs with maximum probability, 0.20. Consecutive missing is favored up to 10 pulses. While ending times of radar sequences are same, beginning times of radar sequences are close. \textbf{Case 4:} No missing occurs. While ending times of radar sequences are different, beginning times of radar sequences might be far. \textbf{Case 5:} Missing occurs with maximum probability, 0.20. Ending and beginning times of radar sequences might be close or far.

In test dataset, we use 15 deinterleaved signals provided by ASELSAN A.Ş. We interleaved random batches of distinct signals while the number of the interleaved signals varies between 2 and 15. The test dataset contains 1000 interleaved sequences and is of different modality and disjoint from training and validation.

\input{figures/pri_fig.tex}

\subsection{Experimental Setup and Evaluation Metrics}

We use Tensorflow \cite{abadi2016tensorflow} deep  learning library throughout the experiments. We set penalty coefficient hyperparameters, $\lambda_{2}{=}10, \lambda_{3}{=}1$ and $\lambda_{4}{=}5$, after fine-tuning the network. Furthermore, we select the maximum of sequence length, the dimension of representation and the number of quantization level as 256, 512 and 5001, respectively. For the optimization procedure, we
choose the base learning rate as $10^{-4}$ and utilize ADAM \cite{kingma2014adam} optimizer for mini-batch gradient descent with a mini-batch size is 128 and default moment parameters. The remained BTE parameters are kept same with \cite{devlin2018bert}.

We use three performance metrics for better analysis: the average probability of the true estimated next PDW position (\text{$Acc_{link}$}), the average probability of the true estimated number of radars (\text{$Acc_{nor}$}) and the average number of one to many assignments (\text{$V_{1{-}m}$}) which violates \eqref{eq:col_const}. The desired values of those metrics are \text{$Acc_{link} {=} 1$}, \text{$Acc_{nor} {=} 1$} and \text{$V_{1-m}{=} 0$}.  Each predicted cluster which has larger than 3 is assigned as a radar signal. 

\subsection{Quantitative Results}
\label{sec:qres}

We examine the effectiveness of the proposed soft min-cost flow (SMCF) framework through evaluation on the validation dataset and disjoint test data. We compare SMCF with CDIF \cite{mardia1989new}, SDIF \cite{milojevic1992improved} and PRIT \cite{nishiguchi2000improved}, which are conventional non-learning based deinterleaving methods, to show their limitations in realistic scenarios. Additionally, we show the significance of relaxed constraints in Eqn. \eqref{eq:final_loss} by training without corresponding penalty terms, \ie $\lambda_i{=}0$, which we denote \textit{Baseline}. We also compare the two inference methods (Greedy, LP) proposed in Sec. \ref{sec:inference}.

Table \ref{tab:results} shows the quantitative results for the deinterleaving tasks. SMCF algorithm consistently outperforms the associated baseline methods by up to 6.1\% and 32.8\% points on \text{$Acc_{link}$} and \text{$Acc_{nor}$} metrics, respectively. Conventional methods perform poorly in challenging scenarios where largely missing pulses and complex PRI modulation types exist. Additionally, the obtained clusters are missing related PDWs although PRI values are predicted correctly.

Comparison of one-to-many violations of Baseline Greedy with of SMCF Greedy as well as the overall performance variation between SMCF Greedy and LP is a good indicator to show the proposed penalty terms improve the solution by enforcing constraints. Similar performances of Greedy\&LP in SMCF and superior performance of Greedy SMCF to Baseline imply constraints are better satisfied in the output of SMCF. We note that all the methods except for Greedy have zero one-to-many violations owing to hard assignment.

\section{Conclusion}
\label{sec:conc}
We link deinterleaving to min-cost flow problem and formulate training of a self-attention based neural network to solve min-cost flow problem. The mechanism behind the formulation is supported by extensive evaluations on the large dataset. Experimental results on data of different modalities show that our framework is able to learn deinterleaving the  signals.

\newpage
\bibliographystyle{IEEEbib}
\bibliography{library}

\end{document}

%% file: packages.tex
\usepackage{import}
\usepackage{spconf}

\usepackage{amsmath}
\usepackage{amsthm}
\usepackage{mathtools}
\usepackage{commath}
\usepackage{mathrsfs}
\usepackage{amssymb}
\usepackage{amsfonts}
\usepackage{bbm}
\usepackage{fancyhdr}
\usepackage{cancel}

\usepackage{algorithm}
\usepackage{algpseudocode}

\usepackage{graphicx}
\usepackage{epsfig}
\usepackage[utf8]{inputenc}
\usepackage{tabularx}
\usepackage{booktabs, multicol, multirow}
\usepackage[table,xcdraw]{xcolor}

\usepackage{caption}
\usepackage{subcaption}

\usepackage{appendix}

\usepackage{cite}
\usepackage{url}

%% file: definitions.tex
\newcommand{\vect}[1]{\boldsymbol{#1}}
\DeclareMathOperator*{\argmin}{arg\,min}

\newcommand{\R}{\ensuremath{\mathbb{R}}}
\newcommand{\T}{\ensuremath{^\intercal}}
\newcommand{\grad}{\ensuremath{\nabla}}

\newcommand{\ie}{\textit{i}.\textit{e}., }

\theoremstyle{definition}

\numberwithin{equation}{section}

%% file: authors.tex
\name{O\u{g}ul Can$^\dagger$ \qquad Yeti Z. G\"{u}rb\"{u}z$^\dagger$ \qquad Berkin Y{\i}ld{\i}r{\i}m$^\ast$ \qquad  A. Ayd{\i}n Alatan$^\dagger$\thanks{This study is supported by ASELSAN A. Ş.}}

\address{$^\dagger$Dept. of Elect. Elec. Eng. \& Center for Image Analysis (OGAM)\\
Middle East Technical University, Ankara, Turkey\\
$^\ast$ASELSAN A. Ş., Ankara, Turkey}

%% file: figures/fig_the_figure_1.tex
\begin{figure}[t]
  \centering
  \centerline{\includegraphics[width=.95\linewidth]{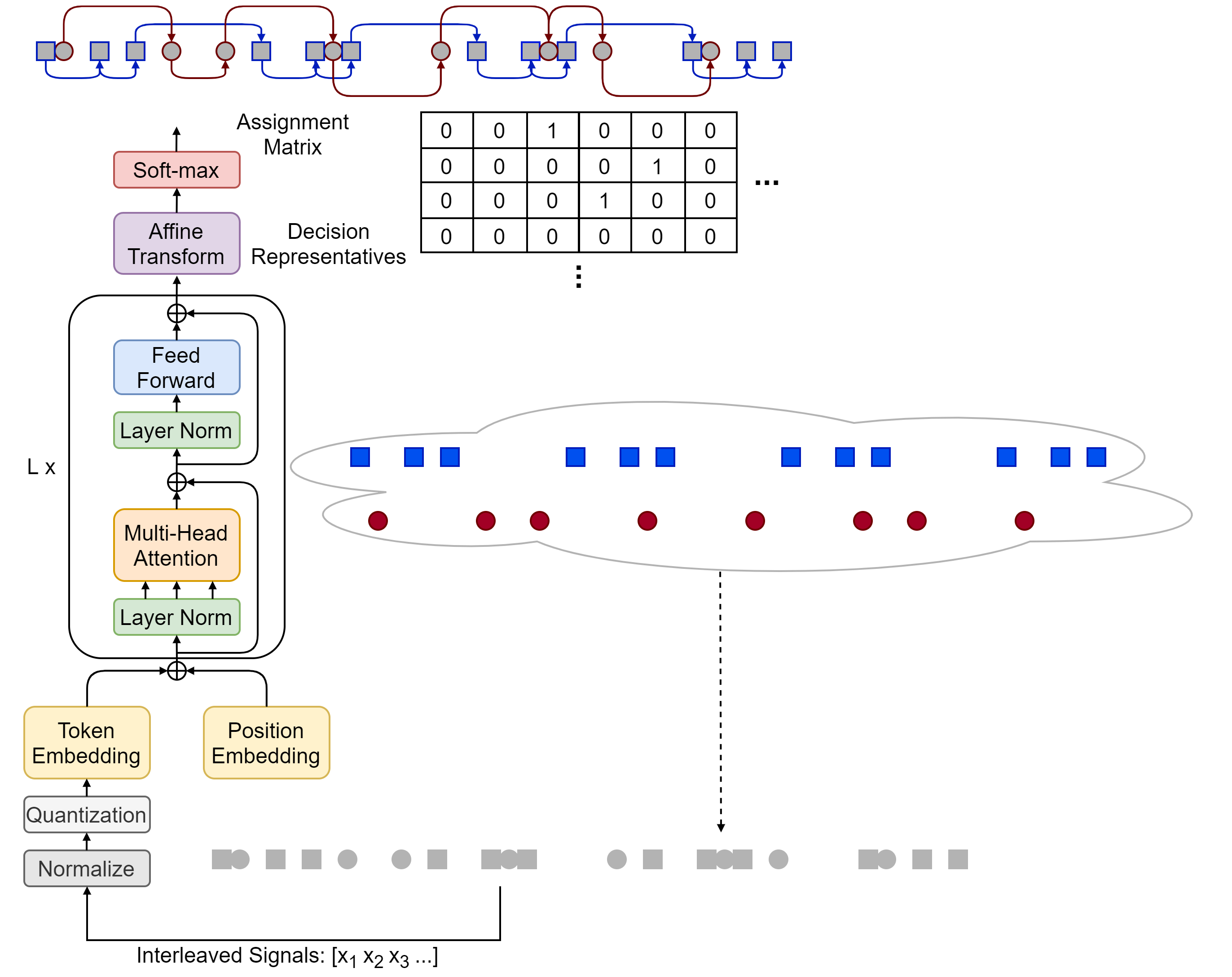}}
  \caption{Overview of the Soft Min-cost Flow Framework}
	\label{fig:the_figure_1}
  \end{figure}

%

%% file: figures/table_performance.tex
\begin{table*}[!t]
\centering
\caption{Comparison with the existing and baseline methods for the deinterleaving task. The best results are indicated in bold.}
\label{tab:results}
\resizebox{\textwidth}{!}{%
\begin{tabular}{c|c|c|c|c|c|c|c|}
Dataset -  \text{$Acc_{link}$} / \text{$Acc_{nor}$} / \text{$V_{1-m}$} & CDIF & SDIF & PRIT & Baseline + Greedy & Baseline + LP & SMCF + Greedy & SMCF + LP \\ \hline
Validation - Case 1 & 0.841 / 0.605 / 0.000 & 0.850 / 0.614 / 0.000 & 0.982 / 0.631 / 0.000 & 0.975 / 0.940 / 0.013 & 0.963 / 0.966 / 0.000 & \textbf{0.991} / \textbf{0.998} / \textbf{0.001} & \textbf{0.991} / \textbf{0.998} / 0.000 \\
Validation - Case 2 & 0.519 / 0.312 / 0.000 & 0.518 / 0.309 / 0.000 & 0.631 / 0.538 / 0.000 & 0.894 / 0.758 / 0.201 & 0.885 / 0.764 / 0.000 & \textbf{0.945} / \textbf{0.996} / \textbf{0.004} & 0.943 / 0.994 / 0.000 \\
Validation - Case 3 & 0.352 / 0.154 / 0.000 & 0.362 / 0.157 / 0.000 & 0.579 / 0.437 / 0.000 & 0.858 / 0.623 / 7.477 & 0.856 / 0.627 / 0.000 & 0.912 / 0.950 / \textbf{0.025} & \textbf{0.915} / \textbf{0.954} / 0.000 \\
Validation - Case 4 & 0.679 / 0.480 / 0.000 & 0.683 / 0.481 / 0.000 & 0.768 / 0.545 / 0.000 & 0.956 / 0.977 / 0.011 & 0.955 / 0.977 / 0.000 & \textbf{0.991} / \textbf{0.999} / \textbf{0.002} & 0.990 / 0.998 / 0.000 \\
Validation - Case 5 & 0.548 / 0.370 / 0.000 & 0.550 / 0.364 / 0.000 & 0.642 / 0.496 / 0.000 & 0.901 / 0.836 / 2.736 & 0.900 / 0.841 / 0.000 & \textbf{0.929} / \textbf{0.975} / \textbf{0.015} & \textbf{0.929 }/ \textbf{0.975} / 0.000 \\
Test & 0.304 / 0.147 / 0.000 & 0.311 / 0.148 / 0.000 & 0.564 / 0.436 / 0.000 & 0.842 / 0.539 / 11.943 & 0.823 / 0.555 / 0.000 & \textbf{0.903} / 0.883 / \textbf{0.098} & 0.898 / \textbf{0.886} / 0.000
\end{tabular}%
}
\end{table*}

%% file: figures/pri_fig.tex
\begin{figure}[t]
  \centering
  \centerline{\includegraphics[width=\linewidth]{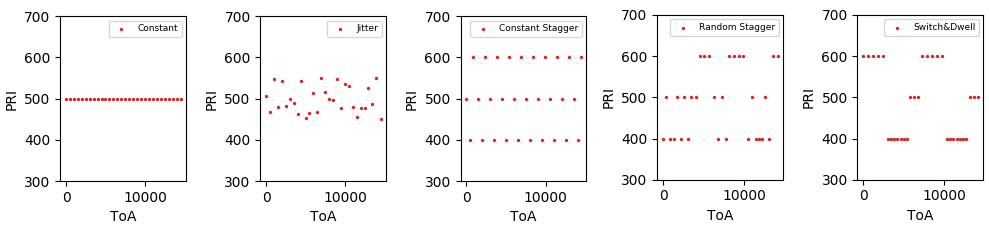}}
  \caption{PRI types. (a) 500 $\mu$s constant PRI. (b) 500 $\mu$s jitter PRI with +/-10\% jitter. (c) [500, 400, 600] $\mu$s constant stagger PRI with 3 stagger levels. (d) [400, 500, 600] $\mu$s random stagger PRI with 3 stagger levels. (e) [600, 400, 500] $\mu$s switch\&dwell PRI with [5, 7, 3] dwells and 3 switches.}
	\label{fig:pri_types}
  \end{figure}

%